\documentclass[12pt]{article}
\usepackage{graphicx}
\usepackage{times}

\usepackage{epsf}
\def\beq{\begin{equation}}
\def\eeq{\end{equation}}
\def\bea{\begin{eqnarray}}
\def\eea{\end{eqnarray}}

\def\vel{\left|}
\def\ver{\right|}
\def\nnb{\nonumber}
\def\ga{\left(}
\def\dr{\right)}

\def\rar{\rightarrow}
\def\nnb{\nonumber}

\def\ba{\begin{array}}
\def\ea{\end{array}}
\def\bea{\begin{eqnarray}}
\def\eea{\end{eqnarray}}

\def\vel{\left|}
\def\ver{\right|}
\def\nnb{\nonumber}
\def\ga{\left(}
\def\dr{\right)}

\def\rar{\rightarrow}
\def\nnb{\nonumber}

\def\lla{\left<}
\def\rra{\right>}

\begin{document}

\title{ {\Large {\bf Semileptonic $\Lambda_b \rar \Lambda \,\nu \bar{\nu}$ decay in
the Leptophobic Z$'$ Model }}}
\author{ {\small B. B.  \c{S}irvanl{\i}}\\
{\small Gazi University, Faculty of Arts and Science, Department of Physics} \\
{\small 06100, Teknikokullar Ankara, Turkey}}

\begin{titlepage}

\maketitle

\begin{abstract}
We study the exclusive flavor changing neutral current process $\Lambda_b
\rar \Lambda \nu \bar{\nu}$ in the leptophobic Z$'$ model, where
charged leptons do not couple to the extra Z$'$ boson. The branching
ratio, as well as, the longitudinal, transversal and normal polarizations
are calculated. It has been shown that all these physical observables
are very sensitive to the existence of new physics beyond the
standard model  and their experimental measurements can give
valuable information about it.
\end{abstract}
~~~PACS number(s):12.60.--i, 13.20.--v, 13.20.He
\end{titlepage}

\section{Introduction \label{s1}}
Flavor-changing neutral current (FCNC) processes present as
stringest tests of the Standart Model (SM) as well as looking for
new physics beyond the SM. In SM, FCNC processes take place only
at the loop level and therefore these decays are suppressed and
for this reason experimentally their measurement is difficult.
Recently, B factory experiments such as Belle and BaBar have
measured FCNC decays due to the $b \rar s \ell^+ \ell^-$
transition at inclusive and exclusive level [1,2,3,4].
Theoretically, the $b \rar s \ell \ell$ transition is described by
the effective Hamiltonian and the FCNC's are represented by QCD
penguin and electroweak penguin (EW) operators. The EW operators
appear in the photon and Z mediated diagrams. Although it is
experimentally very difficult to measure, $B \rar X \,\nu
\bar{\nu}$ decay is an extremely good and "pure" channel to study
Z mediated EW penguin contribution within the SM and its beyond
[5,6,7,8,9].

    Another exclusive decay which is described at inclusive level
by the $b \rar s \ell^+ \ell^-$ transition is the baryonic
$\Lambda_b \rar \Lambda \ell \ell$ decay. Unlike the mesonic
decays, the baryonic decay could maintain the helicity structure
of the effective Hamiltonian for the $b \rar s $ transition [10].

   In this work, we study the $\Lambda_b \rar \Lambda \,\nu
\bar{\nu}$ decay in the leptophobic Z$'$ Model, as a possible
candidate of new physics in the EW penguin sector. The leptophobic
Z$'$ gauge bosons appear naturally in Grand Unified Theories and
string inspired $E_{6}$ models. Here we would like to note that a
similar mesonic $B \rar K(K^{*}) \,\nu \bar{\nu}$ decay and
$\Delta m_{B_{s}}$ mass difference of $B_{s}$ meson in the context
of leptophobic Z$'$ model are studied in [11] and [12],
respectively. In section 2, we briefly consider the leptophobic
model based on $E_{6}$ model. In section 3, we present the
analytical expressions for the branching ratio and the
longitudinal, transverse and normal polarizations of $\Lambda$
baryon, as well as polarization of $\Lambda_b$. In Section 4, we
give numerical analysis and our concluding remarks.

\section{Theoretical background on leptophobic Z$'$ model \label{s2}}

In this section we briefly review the leptophobic Z$'$ model. From
GUT or string-inspired point of view, the $E_{6}$ model [13] is a
very natural extention of the SM. In this model $U(1)^{'}$ gauge
group remains after the symmetry breaking of the $E_{6}$ group. We
assume that the $E_{6}$ group is broken through the following way
$E_{6} \rar SO(10)\times U(1)_{\psi} \rar SU(5)\times
U(1)_{\chi}\times U(1)_{\psi} \rar SU(2)_{L}\times U(1)\times
U(1)^{'}$, where $U(1)^{'}$ is a linear combination of two
additional $U(1)$ gauge groups with  $Q^{'} = Q_{\psi}
\cos\theta-Q_{\chi}\sin\theta$, where $\theta$ is the mixing
angle. When all couplings are GUT normalized, the interaction,
Lagrangian of fermions with Z$'$ gauge boson can be written as
\bea{\cal L}_{int} &=& -\lambda
\frac{g_{2}}{\cos\theta_{w}}\sqrt{\frac{5
\sin^{2}\theta_{w}}{3}}\bar{\psi}\gamma_{\mu}\Bigg(Q^{'}+\sqrt{\frac{3}{5}}
\delta Y_{SM}\Bigg)\psi Z_{\mu}^{'}~, \eea where $\lambda =
\frac{g_{Q^{'}}}{g_{y}}$ is the ratio of gauge couplings $\delta =
- \tan\chi/\lambda$ [14]. In the general case fermion--Z$'$ gauge
boson couplings contain two arbitrary free parameters $\tan\theta$
and $\delta$ [15]. From Eq.(1), it follows that the Z$'$ gauge
boson can be leptophobic when $Q^{'}+\sqrt{\frac{3}{5}} \delta
Y_{SM} = 0$ for the lepton doublet and $e^c$, simultaneously. In
the conventional embedding, the Z$'$ boson can be made leptophobic
with $\delta = -1/3$ and $\tan\theta = \sqrt{\frac{3}{5}}$. In the
Leptophobic Z$'$ model, FCNC's can arise through the mixing
between SM fermions and exotic fermions. In principle, the mixing
of the left handed fermions can lead to the large Z-mediated
FCNC's. In order to forbid this large Z-mediated FCNC's, we
consider the case when mixing take place only between the
right-handed fermions. So, the mixing between right-handed
ordinary and exotic fermions can induce the FCNC's when their Z$'$
charges are different [16]. In the Leptophobic Z$'$ model, the
Lagrangian for $b \rar q (q=s,d)$ transition containing FCNC at
tree level can be written as \bea{\cal L}^{Z^{'}} &=&
-\frac{g_{2}}{2\cos\theta_{w}}U_{qb}^{Z^{'}} \bar
q_{R}\gamma^{\mu}b_{R}Z_{\mu}^{'}~.\eea Using upper experimental
bounds on branching ratio for the $B \rar K \,\nu \bar{\nu}$ [17]
and $B \rar \pi \,\nu \bar{\nu}$ [18] in [11] for model parameters
$U_{sb}^{Z^{'}}$ and $U_{db}^{Z^{'}}$ following bounds are
obtained : $|U_{sb}^{Z^{'}}|\leq0.29$ and
$|U_{db}^{Z^{'}}|\leq0.61$. Analysis of $\Delta m_{B_{s}}$ leads
to the more stringest bound for the $U_{sb}^{Z^{'}}$ [12]:
$|U_{sb}^{Z^{'}}|\leq0.036$ at $m_{Z'}= 700 GeV$ and
$|U_{db}^{Z^{'}}|\leq0.051$ at $m_{Z'}=1 TeV$. These constraints
on $U_{qb}^{Z^{'}}$ we will use in our numerical calculations.

\section{Matrix Elements for the $\Lambda_b \rar \Lambda \,\nu \bar{\nu}$ Decay \label{s3}}

The $\Lambda_b \rar \Lambda \,\nu \bar{\nu}$ decay at quark level
is described by $b \rar s \,\nu \bar{\nu}$ transition. In the SM,
effective Hamiltonian responsible for the $b \rar s \,\nu
\bar{\nu}$ transition is given by

\bea \label{effH} {\cal H}_{eff}^{SM} &=&
\frac{G_{F}\,\alpha}{2\pi\sqrt{2}} V_{tb} V_{ts}^* C_{10} \,\bar
s\gamma_{\mu}(1-\gamma_5)b\,\bar \nu\gamma_\mu(1-\gamma_5)\nu~, \eea
where $G_{F}$ and $\alpha$ are the Fermi and fine structure
constant respectively, $V_{tb}$ and $V_{ts}$ are elements of the
Cabibbo-Kobayashi-Maskawa (CKM) matrix, $C_{10}$ is the Wilson
coefficient whose explicit form can be found in [17,18,19]. As we have
already mentioned, in the leptophobic model, $U(1)^{'}$ charge
is zero for all the ordinary left and right handed lepton fields
within the SM. However, 27 representation of $E_{6}$ contain new
right handed neutrino which is absent in SM and it can give
additional contribution to the processes due to the $b \rar s
\,\nu \bar{\nu}$ transition. Then the effective Hamiltonian
describing $b \rar s \,\nu_{R} \bar{\nu_{R}}$ transition can be
written as \bea \label{effH} {\cal H}_{eff}^{new} &=&
\frac{\pi\,\alpha}{\sin^{2}2\theta_{w}M_{Z^{'}}^{2}}
U_{sb}^{Z^{'}} Q_{\nu_{R}}^{Z^{'}} \,\bar
s\gamma_{\mu}(1+\gamma_5)b\,\bar \nu\gamma_\mu(1+\gamma_5)\nu~, \eea
where $Q^{Z'}_{\nu_{R}}=\frac{1}{2}
x\sqrt{\frac{5\sin^{2}\theta_{w}}{3}}$. $U^{Z'}_{qb}$ and $x$ are
model dependent parameters and we take $x=1$ for simplicity. Having
effective Hamiltonian for $b \rar s \,\nu \bar{\nu}$ transition,
our next problem is the derivation of the decay amplitude for the
$\Lambda_b \rar \Lambda \,\nu \bar{\nu}$ decay, which can be
obtained by calculating the matrix element of the effective
Hamiltonian for $b \rar s \,\nu \bar{\nu}$ transition between
initial $\Lambda_b$ and final $\Lambda$ baryon states. It follows
that the matrix elements $\lla \Lambda \vel \bar s \gamma_\mu
(1\mp\gamma_{5}) b \ver \Lambda_b \rra$ are needed for calculating
the decay amplitude for the $\Lambda_b \rar \Lambda \,\nu
\bar{\nu}$ decay. These matrix elements parametrized in terms of
the form factors are as follows ([20],[21])

 \bea \label{mel1} \lla
\Lambda \vel \bar s \gamma_\mu b \ver \Lambda_b \rra &=& \bar
u_\Lambda \Bigg[f_1 \gamma_\mu + i\,f_2 \sigma_{\mu\nu} q^{\nu} +
f_3 q_{\mu}\bar \Bigg]u_{\Lambda_{b}}~, \eea

\bea \label{mel2} \lla \Lambda \vel \bar s \gamma_\mu\gamma_5 b
\ver \Lambda_b \rra &=& \bar u_\Lambda\Bigg[g_1 \gamma_\mu\gamma_5
+ i\,g_2 \sigma_{\mu\nu} q^{\nu}\gamma_5 + g_3 q_{\mu}\gamma_5
\Bigg] u_{\Lambda_{b}} \eea where $q=p_{\Lambda_{b}}-p_{\Lambda}$
is the momentum transfer and $f_i$ and $g_i$ $(i=1,2,3)$ are the
form factors. Note that form factors $f_3$ and $g_3$ do not give
contribution to the considered decay since neutrinos are
massless. Using Eqs. (3)-(4)-(6) and (7) for the decay amplitude of
the $\Lambda_b \rar \Lambda \,\nu \bar{\nu}$ decay in leptophobic
model, we have \bea \label{ib} {\cal M}&=& \frac{\alpha G_F}{2
\sqrt{2} \, \pi} V_{tb} V_{ts}^* C_{10}^{\nu} \Bigg\{\bar
\nu\gamma_\mu(1-\gamma_5)\nu \times \bar u_\Lambda \Bigg[A_1
\gamma_\mu(1+\gamma_5)
+ B_1 \gamma_\mu(1-\gamma_5) \nnb \\
&+&i\,\sigma_{\mu\nu}q^{\nu}A_2(1+\gamma_5)+ B_2(1-\gamma_5))
\Bigg]u_{\Lambda_{b}}\nnb
\\&+&C_{RR}\bar \nu\gamma_\mu(1+\gamma_5)\nu \times \bar u_\Lambda
\Bigg[B_1\gamma_\mu(1+\gamma_5)
+A_1\gamma_\mu(1-\gamma_5) \nnb \\
&+&i\,\sigma_{\mu\nu}q^{\nu}B_2(1+\gamma_5)+
A_2(1-\gamma_5))\Bigg]u_{\Lambda_{b}}\Bigg\}~, \eea where
$C_{RR}=\frac{\pi Q U_{qb}^{Z'}}{\alpha V_{tb} V_{ts}^*
C_{10}^{\nu}}\frac{m_{Z}^{2}}{m_{Z'}^{2}}$.

It is shown in [21] that when HQET is applied the number of
independent form factors are reduced to two ($F_{1}$ and $F_{2}$),
irrelevant of the Dirac structure of the corresponding operators,
i.e., $\lla\Lambda(p_{\Lambda})\vel\bar s\Gamma b\ver
\Lambda(p_{\Lambda_{b}})\rra = \bar u_\Lambda\Bigg[F_{1}(q^2) +
{\not\!\upsilon}F_{2}(q^2)\Bigg]\Gamma u_{\Lambda_{b}}$, where
$\Gamma$ is an arbitrary Dirac matrix and
$\upsilon^{\mu}=p^{\mu}/m_{\Lambda_{b}}$ is the four velocity of
$\Lambda_{b}$. Using this result we get \bea
f_1&=&g_1=F_1+\sqrt{r}F_2,\nnb \\
f_2&=&g_2=\frac{F_2}{m_{\Lambda_{b}}}~,\eea where $r =
m_{\Lambda}^{2}/m_{\Lambda_{b}}^{2}$.

Our next problem is the calculation of $\Lambda$ baryon
polarizations using the matrix element in Eq.(7). In the rest
frame of $\Lambda$, the unit vectors along the longitudinal,
normal and transversal components of $\Lambda$ are chosen as \bea
s_L^\mu &=& \ga 0,\vec{e}_L \dr = \left( 0, \frac{\vec{p}_\Lambda}
{\vel \vec{p}_\Lambda \ver} \right)~, \nnb \\
s_T^\mu &=& \ga 0,\vec{e}_T \dr =   \left( 0, \frac{\vec{e}_L \times
\vec{\xi}_{\Lambda_b}}{\vel \vec{e}_L \times \vec{\xi}_{\Lambda_b}
\ver} \right)~, \nnb \\
s_N^\mu &=& \ga 0,\vec{e}_N \dr = \left( 0,\vec{e}_L \times \vec{e}_T
\right)~.
\eea
The longitudinal component of the $\Lambda$ baryon polarization is boosted
to the center of mass frame of the nuetrino--antineutrino pair by Lorentz
transformation, yielding
\bea
\ga s_L^\mu \dr_{CM} = \ga \frac{\vel \vec{p}_\Lambda \ver}{m_\Lambda},
\frac{E_\Lambda \vec{p}_\Lambda }{m_\Lambda \vel \vec{p}_\Lambda \ver} \dr~,
\eea
where $E_\Lambda$ and $m_\Lambda$ are the energy and mass of the $\Lambda$
baryon in the CM frame of nuetrino--antineutrino pair. The remaining two
unit vectors $s_T^\mu$ and $s_N^\mu$ are unchanged under Lorentz
transformation.

After integration over neutrino and antineutrino momentum the
differential decay rate of the $\Lambda_b \rar \Lambda \,\nu
\bar{\nu}$ decay for any spin direction $\vec{\xi}$ along the
$\Lambda$ baryon can be written as (in the rest frame of
$\Lambda_b$) \bea d\Gamma=\frac{1}{4}\,\,d\Gamma^{0}
\Bigg[1+\frac{I_{1}}{I_{0}}\vec{e}_{L}\cdot\vec{\xi}_{\Lambda_{b}}\Bigg]
\Bigg[1+\vec{P}_{\Lambda}\cdot \vec{\xi_{\Lambda}}\Bigg].\eea
\bea\vec{P}_{\Lambda}=\frac{1}{1+\frac{I_{1}}{I_{0}}\hat{e}_{L}\cdot
\hat\xi_{\Lambda_{b}}}\Bigg[\Bigg(\frac{I_2}{I_0}+\frac{I_3}{I_0}\hat{e}_{L}\cdot
\hat\xi_{\Lambda_{b}}\Bigg)\hat{e}_{L}+\frac{I_4}{I_0}\hat{e}_{N}
+\frac{I_5}{I_0}\hat{e}_{T}\Bigg]\eea \bea
d\Gamma^{0}=\frac{3G_{F}^{2}\alpha_{em}^{2}|V_{tb} V_{ts}^*|^2
|C_{10}|^2}{32.32.4\pi^{7}m_{\Lambda_{b}}}|\vec{p_{\Lambda}}|I_0
dE_{\Lambda} d\Omega_{\Lambda}\eea The functions $I_{i}$ in
Eq.(13) have following expressions: \bea \label{Delta} I_{0}&=&
\frac{32}{3}\pi\Bigg\{ (q^2p_{\Lambda}\cdot
p_{\Lambda_{b}}+2p_{\Lambda}\cdot q\,p_{\Lambda_{b}}\cdot q)
(|A_1|^2+|B_1|^2+|C_{RR}|^2(|B_1|^2+|A_1|^2))\nnb\\&+&
(q^2p_{\Lambda}\cdot p_{\Lambda_{b}}-4p_{\Lambda}\cdot
q\,p_{\Lambda_{b}}\cdot q)q^2
(|A_2|^2+|B_2|^2+|C_{RR}|^2(|B_2|^2+|A_2|^2))\nnb\\&+&
6m_{\Lambda}q^2 p_{\Lambda_{b}}\cdot q
\Bigg[Re[A_1A_2^{\ast}]+Re[B_1B_2^{\ast}]+|C_{RR}|^2Re[B_1B_2^{\ast}]
+|C_{RR}|^2Re[A_1A_2^{\ast}]\Bigg] \nnb\\&-&6m_{\Lambda_{b}}q^2
p_{\Lambda}\cdot q
\Bigg[Re[A_2B_1^{\ast}]+Re[A_1B_2^{\ast}]+|C_{RR}|^2Re[B_2A_1^{\ast}]
+|C_{RR}|^2Re[B_1A_2^{\ast}]\Bigg]
\nnb\\&-&6m_{\Lambda_{b}}m_{\Lambda}q^2
\Bigg[Re[A_1B_1^{\ast}]+q^2Re[A_2B_2^{\ast}]+|C_{RR}|^2Re[B_1A_1^{\ast}]\nnb\\&+&
|C_{RR}|^2q^2Re[B_2A_2^{\ast}]\Bigg] \Bigg\}\eea

\bea \label{PLI1} I_{1}&=&\frac{32}{3I_{0}}\pi p_{\Lambda}\Bigg\{
(-m_{\Lambda_{b}}) (q^2-2p_{\Lambda}\cdot
q)\Bigg[(|A_1|^2-|B_1|^2)+|C_{RR}|^2(|B_1|^2-|A_1|^2)\Bigg]\nnb\\&+&
m_{\Lambda_{b}}q^2(q^2+4p_{\Lambda}\cdot
q)\Bigg[(|A_2|^2-|B_2|^2)+|C_{RR}|^2(|B_2|^2-|A_2|^2)\Bigg]\nnb\\&+&
6m_{\Lambda_{b}}m_{\Lambda}q^2
\Bigg[Re[A_1A_2^{\ast}]-Re[B_1B_2^{\ast}]+|C_{RR}|^2Re[B_1B_2^{\ast}]
-|C_{RR}|^2Re[A_1A_2^{\ast}]\Bigg]\nnb\\&+&
2q^2(p_{\Lambda_{b}}\cdot p_{\Lambda}+p_{\Lambda_{b}}\cdot
q)\Bigg[Re[A_1B_2^{\ast}]-Re[A_2B_1^{\ast}]-|C_{RR}|^2Re[B_2A_1^{\ast}]\nnb\\&+&
|C_{RR}|^2Re[B_1A_2^{\ast}]\Bigg]\Bigg\}\eea

\bea \label{PLI2} I_{2}&=&\frac{32}{3I_{0}}\pi
p_{\Lambda}m_{\Lambda_{b}}\Bigg\{2m_{\Lambda}p_{\Lambda}(|A_1|^2+|B_1|^2)+(q^2+
p_{\Lambda_{b}}\cdot q)(|A_1|^2-|B_1|^2)
\nnb\\&+&2q^2(3m_{\Lambda_{b}}+p_{\Lambda})
\Bigg[Re[A_2B_1^{\ast}]+|C_{RR}|^2Re[B_2A_1^{\ast}]\Bigg]\nnb\\&-&
2q^2(3m_{\Lambda_{b}}-p_{\Lambda})
\Bigg[Re[A_1B_2^{\ast}]+|C_{RR}|^2Re[B_1A_2^{\ast}]\Bigg]\nnb\\&-&
\frac{2}{m_{\Lambda}}p_{\Lambda}q^4\Bigg[Re[A_2B_2^{\ast}]+|C_{RR}|^2Re[B_2A_2^{\ast}]\Bigg]
\nnb\\&+&\frac{2}{m_{\Lambda}}q^2(m_{\Lambda}p_{\Lambda}+p_{\Lambda_{b}}\cdot
p_{\Lambda}+p_{\Lambda}\cdot
q)\Bigg[Re[B_1B_2^{\ast}]+|C_{RR}|^2Re[A_1A_2^{\ast}]\Bigg]
\nnb\\&+&\frac{2}{m_{\Lambda}}q^2(m_{\Lambda}p_{\Lambda}-p_{\Lambda_{b}}\cdot
p_{\Lambda}+p_{\Lambda}\cdot
q)\Bigg[Re[A_1A_2^{\ast}]+|C_{RR}|^2Re[B_1B_2^{\ast}]\Bigg]
\nnb\\&+&\frac{2p_{\Lambda}}{m_{\Lambda}}(q^2-2p_{\Lambda_{b}}\cdot
p_{\Lambda}+2p_{\Lambda}\cdot q-2p_{\Lambda_{b}}\cdot q)
\Bigg[Re[A_1B_1^{\ast}]+|C_{RR}|^2Re[B_1A_1^{\ast}]\Bigg]\nnb\\&+&
|C_{RR}|^2(2m_{\Lambda_{b}}p_{\Lambda}-q^2-2p_{\Lambda_{b}}\cdot
q)\Bigg[|E_{1}|^2+|D_{1}|^2\Bigg]\nnb\\&-& q^2
(4m_{\Lambda_{b}}p_{\Lambda}+q^2-4p_{\Lambda_{b}}\cdot
q)\Bigg[|A_{2}|^2+|B_{2}|^2\Bigg] \nnb\\&-& q^2|C_{RR}|^2
(4m_{\Lambda_{b}}p_{\Lambda}+q^2-4p_{\Lambda_{b}}\cdot
q)\Bigg[|D_{2}|^2+|E_{2}|^2\Bigg] \Bigg\}\eea

\bea \label{PLI3} I_{3}&=&\frac{64\pi}{3I_{0}}
\Bigg\{\frac{E_{\Lambda}}{m_{\Lambda}}\Bigg[q^4p_{\Lambda}p_{\Lambda_{b}}\Bigg(
Re[A_2B_2^{\ast}]-|C_{RR}|^2(Re[B_1A_1^{\ast}]-Re[B_2A_2^{\ast}])\Bigg)
\nnb\\&-&m_{\Lambda_{b}}q^2p_{\Lambda}\cdot
q\Bigg[Re[A_1A_2^{\ast}]+Re[B_1B_2^{\ast}]-|C_{RR}|^2(Re[B_1B_2^{\ast}]+
Re[A_1A_2^{\ast}])\Bigg]\Bigg]\nnb\\&+&
m_{\Lambda}q^2p_{\Lambda_{b}}\cdot
q\Bigg[Re[A_2B_1^{\ast}]+Re[A_1B_2^{\ast}]+|C_{RR}|^2(Re[B_2A_1^{\ast}]+
Re[B_1A_2^{\ast}])\Bigg]\nnb\\&+&|C_{RR}|^2p_{\Lambda_{b}}\cdot q
p_{\Lambda}\cdot q Re[B_1A_1^{\ast}]-(q^2p_{\Lambda_{b}}\cdot
p_{\Lambda}-2p_{\Lambda_{b}}\cdot q p_{\Lambda}\cdot
q)Re[A_1B_1^{\ast}]\Bigg]\nnb\\&+&\frac{m_{\Lambda_{b}}}
{m_{\Lambda}}p_{\Lambda}^{2}\Bigg[m_{\Lambda_{b}}q^2\Bigg(
Re[A_1A_2^{\ast}]+Re[B_1B_2^{\ast}]+|C_{RR}|^2
(Re[B_1B_2^{\ast}]+Re[A_1A_2^{\ast}])\Bigg)\nnb\\&+&m_{\Lambda}q^2\Bigg(
Re[A_2B_1^{\ast}]+Re[A_1B_2^{\ast}]+|C_{RR}|^2
(Re[B_2A_1^{\ast}]+Re[B_1A_2^{\ast}])\Bigg)\Bigg] \Bigg\}\eea

\bea \label{PN} I_{4}&=&\frac{16\pi}{3I_{0}}\Bigg\{
-m_{\Lambda}m_{\Lambda_{b}}q^2\Bigg[(|A_1|^2+|B_1|^2)+q^2(|A_2|^2+|B_2|^2)
+|C_{RR}|^2(|B_1|^2+|A_1|^2)\nnb\\&+&|C_{RR}|^2q^2(|B_2|^2+|A_2|^2)\Bigg]+
2q^4p_{\Lambda}\cdot p_{\Lambda_{b}}
\Bigg[Re[A_2B_2^{\ast}]+|C_{RR}|^2Re[B_2A_2^{\ast}]\Bigg]\nnb\\&-&
2m_{\Lambda_{b}}q^2p_{\Lambda}\cdot q
\Bigg[Re[A_1A_2^{\ast}]+Re[B_1B_2^{\ast}]+|C_{RR}|^2(Re[B_1B_2^{\ast}]
+Re[A_1A_2^{\ast}])\Bigg]\nnb\\&+& 2m_{\Lambda}q^2p_{\Lambda}\cdot
q\Bigg[Re[A_2B_1^{\ast}]+Re[A_1B_2^{\ast}]+|C_{RR}|^2(Re[B_2A_1^{\ast}]
+Re[B_1A_2^{\ast}])\Bigg]\nnb\\&-& 2(q^2p_{\Lambda}\cdot
p_{\Lambda_{b}}-2p_{\Lambda}\cdot q\,p_{\Lambda_{b}}\cdot q)
\Bigg[Re[A_1B_1^{\ast}]+|C_{RR}|^2Re[B_1A_1^{\ast}]\Bigg]
 \Bigg\}\eea

 \bea \label{PT}
I_{5}&=&\frac{16\pi m_{\Lambda_{b}}p_{\Lambda}}{3I_{0}}\Bigg\{
2m_{\Lambda_{b}}q^2\Bigg[Im[A_1A_2^{\ast}]+Im[B_1B_2^{\ast}]+
|C_{RR}|^2(Im[B_1B_2^{\ast}]+Im[A_1A_2^{\ast}])\Bigg]\nnb\\&+&
2m_{\Lambda}q^2\Bigg[Im[A_2B_1^{\ast}]+Im[A_1B_2^{\ast}]+
|C_{RR}|^2(Im[B_2A_1^{\ast}]+Im[B_1A_2^{\ast}])\Bigg]\nnb\\&-&
2q^4\Bigg[Im[A_2B_2^{\ast}]+|C_{RR}|^2Im[B_2A_2^{\ast}]\Bigg]\nnb\\&+&
2(2m_{\Lambda_{b}}^{2}-q^2-2p_{\Lambda}\cdot p_{\Lambda_{b}})
\Bigg[Im[A_1B_1^{\ast}]+|C_{RR}|^2Im[B_1A_1^{\ast}]\Bigg]
\Bigg\}\eea

Here the kinematics and the relationships for the form factors are
given as follows

\bea
q^2&=&m_{\Lambda_{b}}^{2}+m_{\Lambda}^{2}-2m_{\Lambda_{b}}E_{\Lambda}\nnb
\\p_{\Lambda}\cdot p_{\Lambda_{b}}&=&m_{\Lambda_{b}}E_{\Lambda}\nnb
\\p_{\Lambda_{b}}\cdot
q&=&m_{\Lambda_{b}}^{2}-m_{\Lambda_{b}}E_{\Lambda}\nnb
\\p_{\Lambda}\cdot q&=&m_{\Lambda_{b}}E_{\Lambda}-m_{\Lambda}^{2}
\eea

and \bea
  A_1&=&(f_1-g_1)\,\,\, , A_2=(f_2-g_2)\,\,\, ,\nnb
\\B_1&=&(f_1+g_1)\,\,\, , B_2=(f_2+g_2)\,\,\,.\eea

\section{Numerical analysis and discussion \label{s4}}
In this section, we present our numerical analysis on the
branching ratio and $P_L,P_N,P_T$ polarizations of $\Lambda$
baryon. The value of input parameters, we use in our calculations
are
\begin{eqnarray} & &
m_{\Lambda_{b}} =5.62\, GeV \, , \, m_{\Lambda} =1.116 \, GeV \, ,  \nnb \\
& & f_B=0.2 \, GeV \, , \, \,\, |V_{tb} V^*_{ts}|=0.04 \, \, , \,
\, \alpha^{-1}=137\, , \nnb \\  & & |C_{10}^{\nu}|=4.6 \,  ,\,\,
G_F=1.17 \times 10^{-5}\, GeV^{-2} \,  ,\,\, \tau_{\Lambda_{b}}=1.24 \times 10^{-12} \, s \,\, , \nnb \\
& & |U_{sb}^{Z'}|\leq0.29\,\, , \,\,\,\,|U_{db}^{Z'}|\leq0.61.
\end{eqnarray} We have assumed that all the neutrinos are
massless. From the expressions of branching ratio and $\Lambda$
baryon polarizations it follows that the form factors are the main
input parameters. The exact calculation for the all form factors
which appear in $\Lambda_{b}\rar \Lambda$ transition does not
exist at present. For the form factors, we will use results from
QCD sum rules method in corporation with heavy quark effective
theory [23,24]. The $q^2$ dependence of the two form factors are
given as follows :\bea F(q^2)=\frac{F(0)}{1+a_{F}
(q^2/m_{\Lambda_{b}}^2) + b_{F} (q^2/m_{\Lambda_{b}}^2)^2}~.\eea The
values of the form factors parameters are given listed in Table1
\begin{table}
        \begin{center}
        \begin{tabular}{|c|c|c|c|c|c|c|c|c|}
        \hline
        \multicolumn{1}{|c|}{ $ $}       &
        \multicolumn{1}{|c|}{ $F(0)$}       &
        \multicolumn{1}{|c|}{ $a_F$}       &
        \multicolumn{1}{|c|}{ $b_F$}         \\
        \hline
        $F_1$ & $0.462$ & $-0.0182$ & $-0.000176$       \\
        \hline
        $F_2$ & $-0.077$ & $-0.0685$ & $0.00146$       \\
        \hline
        \end{tabular}
        \end{center}
\caption{ Form factors for $\Lambda_b \rar \Lambda \,\nu
\bar{\nu}$ decay in a three parameters fit.\label{table1}}
\end{table}.

Now, let us examine the new effects to the branching ratio of the
$\Lambda_b \rar \Lambda \, \nu \bar{\nu}$ decay and polarization
effects of the $\Lambda$ baryon in the leptophobic Z$'$ boson. In
the Leptophobic Z$'$ model,there are two new parameters, namely,
the effective FCNC coupling constant $|U_{sb}^{Z'}|$ and the mass
for the Z$'$ boson. Although the mass of Z$'$ boson range is given
$365\,GeV\,\leq M_{Z'}\leq615\,GeV$ in the D0 experiment, we take
$M_{Z'}=700GeV$ and $|U_{sb}^{Z'}|\leq0.29$ [22], which we will
use in our calculations. In Fig. 1, we present the dependence of
the differential decay branching ratio (BR), for $\Lambda_b \rar
\Lambda \,\nu \bar{\nu}$ decay, as function of the momentum
square,$q^2$. We observe from this figure that the differential
branching ratio at low $q^2$ region is one order larger compared
to that of the SM prediction. In Fig.2, we show our prediction for
the branching ratio as a function of the effective coupling
constant $|U_{sb}|$ in the Leptophobic Z$'$ Model. We see from
this figure with increasing $|U_{sb}|$, BR also increases. We see
that difference between two models is clear when $|U_{sb}|\geq
0.4$. In Fig.3, the dependence of the longitudinal polarization
$P_L$ for $\Lambda_b \rar \Lambda \,\nu \bar{\nu}$ decay is
presented. For completeness, in this figure we also present the
prediction of SM for $P_L$. From this figure we see that the
magnitude and sign of $P_L$ are different in these models.
Therefore a measurement of the magnitude and sign of $P_L$ can
provide valuable information about the new physics effects. With
increasing $q^2$ the difference between two models decreases. At
least up to $q^2=10 GeV^{2}$ there is noticeable difference
between these models and this can serve as a good test for
discrimination of these theories. The transverse component $P_T$
of the $\Lambda$ polarization is a T-odd quantity. A nonzero value
of $P_T$ could indicate CP violation. In the SM, there is no CP
violating phase in the CKM element of $V_{tb} V^*_{ts}$ and since
parametrization of the form factors are real, they can not induce
$P_T$ in the $\Lambda_b \rar \Lambda \,\nu \bar{\nu}$ decay.
Therefore if transverse $\Lambda$ polarization is measured in the
experiments to be nonzero, it is an indication of the existence of
new CP violating source and new types of interactions. The
parameter $U_{sb}^{Z'}$ in leptophobic Z$'$ model, in principle,
can have imaginary part and therefore it can induce CP violation.
But in the considered model, there is no interference terms
between SM and leptophobic Z$'$ model contributions and terms
proportional to $U_{sb}^{Z'}$ involves to the branching ratio and
polarization effects in the form $|U_{sb}^{Z'}|^2$. Therefore it
cannot induce CP violation. Depicted in Fig.4 is the dependence of
the normal polarization $P_N$ for $\Lambda_b \rar \Lambda \,\nu
\bar{\nu}$ decay. The difference between the two models in
prediction of $P_N$ becomes considerable at $q^2\geq10 GeV^{2}$.
Therefore analysis of $P_N$ in this region can be informative for
discriminating these models.

When $\Lambda$ is not polarized, summing over the $\Lambda$ spin in Eq.(11),
we get \bea
d\Gamma=\frac{d\Gamma^{0}}{2}\Bigg(1+
\alpha_{\Lambda_{b}}\hat{e}_{L}\cdot \hat
\xi_{\Lambda_{b}}\Bigg)~,\eea where
$\alpha_{\Lambda_{b}}=\frac{I_1}{I_0}$. So, the polarization of
$\Lambda_{b}$ is $P_{\Lambda_{b}}\equiv \alpha_{\Lambda_{b}}$. In
Fig.5 we present the dependence of polarization of $\Lambda_{b}$
baryon $\alpha_{\Lambda_{b}}$ on $q^2$. We see that up to $q^2=13
GeV^2$ the sign of $\alpha_{\Lambda_{b}}$ in leptophobic model is
positive and negative in SM. When $q^2>13 GeV^2$ situation becomes
vice versa. For this reason measurement of the sign and magnitude of
$\alpha_{\Lambda_{b}}$ at different $q^2$ can provide us essential
information about existence of new physics. For unpolarized
$\Lambda_{b}$ ; i.e. $\hat \xi_{\Lambda_{b}}=0$, we have
\bea\vec{P_{\Lambda}}=\alpha_{\Lambda}\hat{e}_{L}~,\eea with
$\alpha_{\Lambda}=\frac{I_2}{I_0}$ which leads to the result that the $\Lambda$
polarization is purely longitudinal, i.e. $P_{\Lambda}=
\alpha_{\Lambda}$ and $P_N=P_T=0$.

In conclusion, we have studied the Semileptonic $\Lambda_b \rar
\Lambda \,\nu \bar{\nu}$ decay in the Leptophobic Z$'$ Model. We
have analysed the longitudinal, transverse and normal polarization
of the $\Lambda$ baryon on $q^2$ dependence. The sensitivity of
the branching ratio on $|U_{qb}^{Z'}|$ is investigated and the
dependence of the polarization parameter of $\Lambda_{b}$ baryon
on $q^2$ is also investigated. We find that all physical
observables are very sensitive to the existence of new physics
beyond SM.

\vspace{2cm}

We would like to thank T. M. Aliev and M. Savc{\i} for useful
discussions.

\newpage
\renewcommand{\topfraction}{.99}
\renewcommand{\bottomfraction}{.99}
\renewcommand{\textfraction}{.01}
\renewcommand{\floatpagefraction}{.99}

\begin{figure}
\centering
\includegraphics[width=5in]{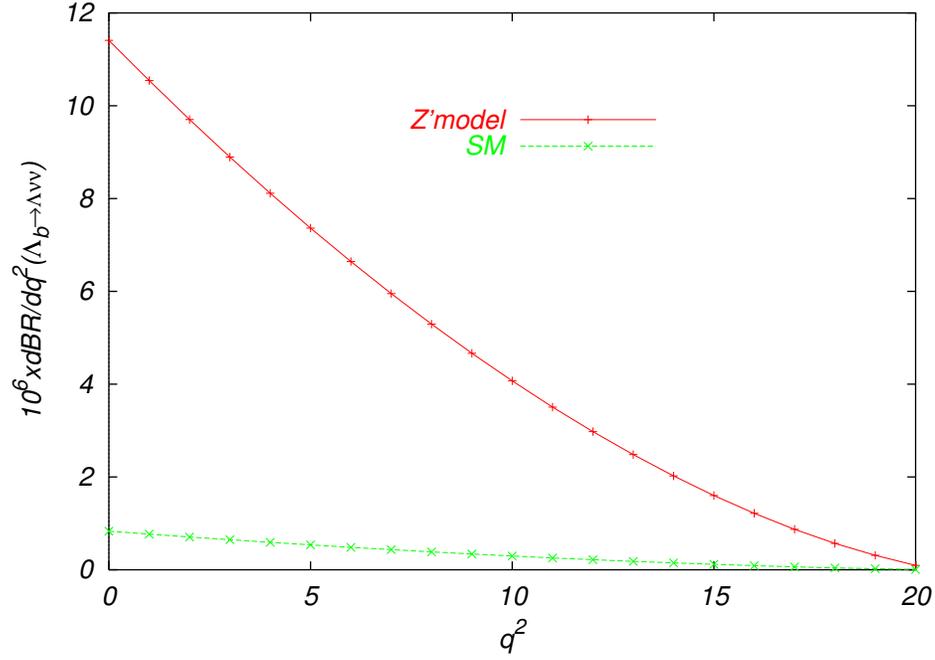}
\caption{The dependence of the differential decay branching
ratio(BR),for $\Lambda_b \rar \Lambda \,\nu \bar{\nu}$ decay, as
function of the momentum transfer,$q^2$ \label{f1}.}
\end{figure}
\begin{figure}
\centering
\includegraphics[width=5in]{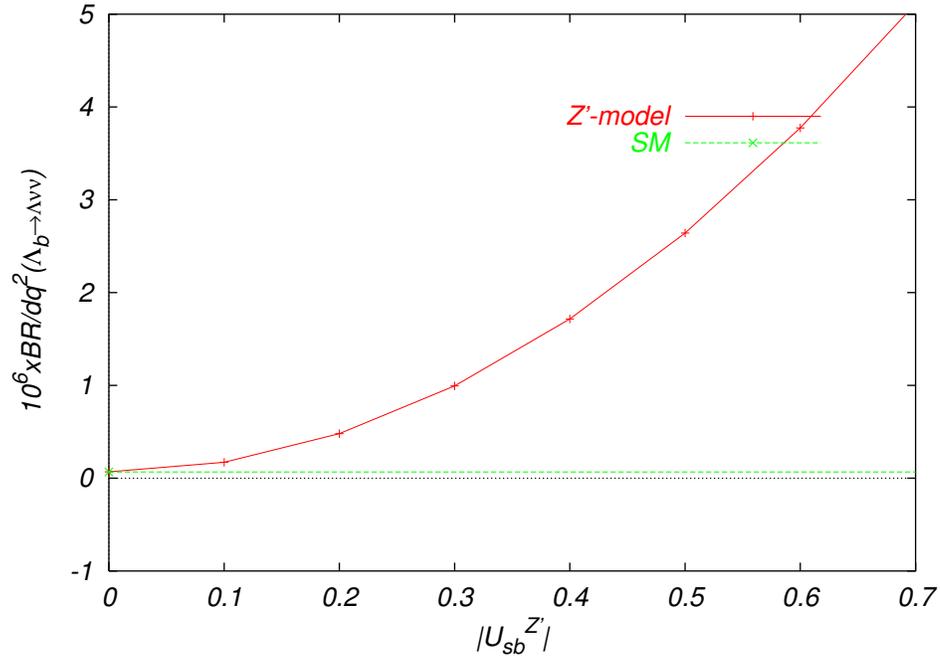}
\caption{Integrated branching ratio(BR)as function of the
$|U_{sb}|$, effective coupling constant in the Leptophobic Z$'$
Model. \label{f2}.}
\end{figure}
\clearpage
\begin{figure}
\centering
\includegraphics[width=5in]{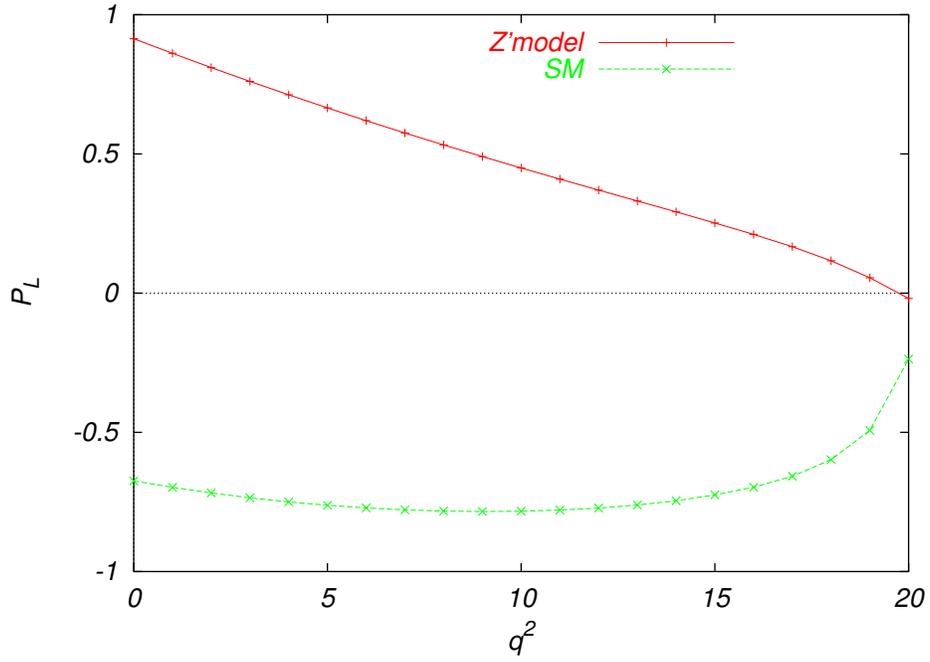}
\caption{The dependence of the longitudinal  polarization $P_L$
for $\Lambda_b \rar \Lambda \,\nu \bar{\nu}$ decay.\label{f3}}
\end{figure}
\begin{figure}
\centering
\includegraphics[width=5in]{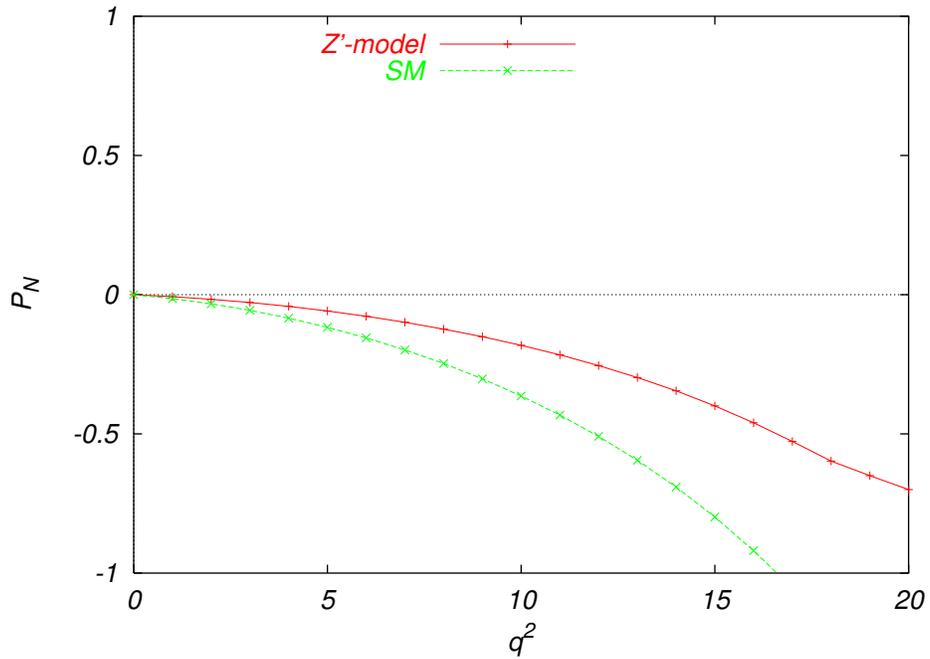}
\caption{The dependence of the normal  polarization $P_N$ for
$\Lambda_b \rar \Lambda \,\nu \bar{\nu}$ decay. \label{f4}}
\end{figure}

\begin{figure}
\centering
\includegraphics[width=5in]{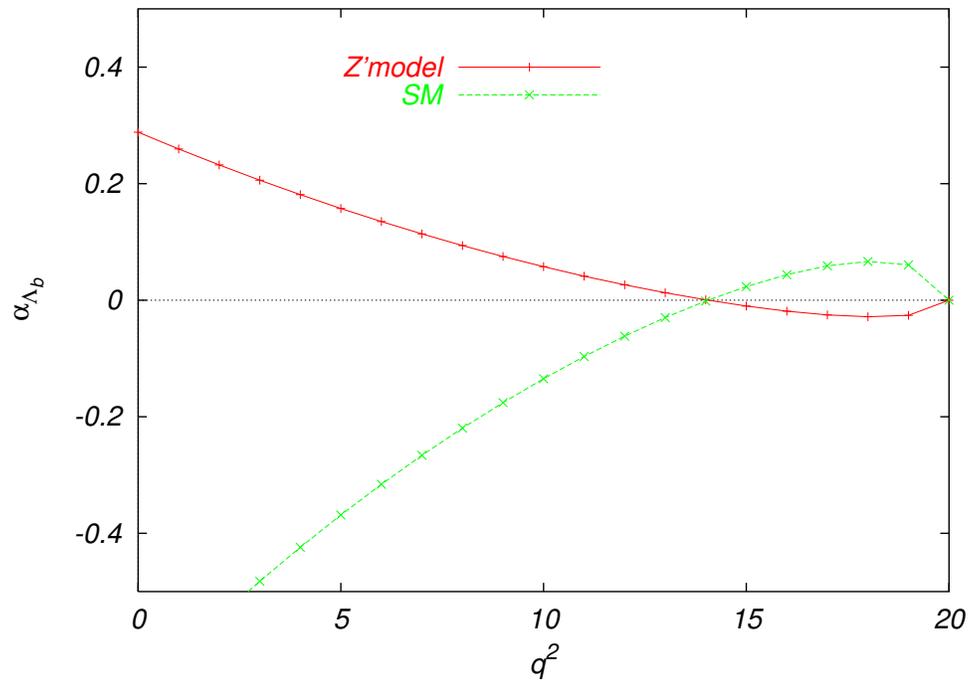}
\caption{The distribution of $\alpha_{\Lambda_{b}}$ as a function
of $q^2$. \label{f5}}
\end{figure}

\clearpage
%

%
%
\end{document}